\documentclass{svmult}

\usepackage[bottom]{footmisc}

\usepackage{amsfonts,amsmath}

\newcommand{\ck}{\mathcal{L}}

\newcommand{\spa}[1]{\mathcal{#1}}

\newcommand{\Rn}{\mathbb{R}^n}

\newcommand{\Rt}{\mathbb{R}^3}

\newcommand{\B}{\mathbf{B}}

\newcommand{\N}{\mathcal{N}}

\newcommand{\Ra}{\mathcal{R}}

\newcommand{\Rc}{\mathcal{R}^{\perp}}

\newcommand{\ek}{\mathcal{E}}

\makeindex             

\begin{document}

\title*{Elliptic systems}
\author{Sergio Dain}
\institute{Max-Planck-Institut f\"ur Gravitationsphysik\\
  Am M\"uhlenberg 1\\
  14476 Golm\\
  Germany\\
\texttt{dain@aei.mpg.de}}
%
%
\maketitle

\begin{abstract}
  In this article I will review some basic results on elliptic
  boundary value problems with applications to General Relativity.
\end{abstract}

\section{Introduction}
\label{sec:in}

Elliptic problems appear naturally in physics mainly in two
situations: as equations which describe equilibrium (for example,
stationary solutions in General Relativity) and as constraints for the
evolutions equations (for example, constraint equations in
Electromagnetism and General Relativity). In addition, in General
Relativity they appear often as gauge conditions for the evolutions
equations.

The model for all elliptic equations is the Laplace equation. Let us 
consider the Dirichlet boundary value problem for this equation 
\begin{equation}
\label{eq:3}
  \Delta u = f\text{ on } \Omega, \quad 
u = g \text{ on } \partial \Omega, 
\end{equation}
where $\Omega$ is a bounded, smooth, domain in $\Rn$ with
boundary $\partial \Omega$;  $f,g$ are  smooth functions and $\Delta$
is the  Laplace operator in $\Rn$. 

It is a well known result that for every source $f$ and every boundary value
$g$ there exist a unique, smooth, solution
$u$ of \eqref{eq:3}.  We would like to generalize equations
\eqref{eq:3} for more general operators
and more general boundary conditions.

The first step in this generalization is given by the Neumann problem 
\begin{equation}
  \label{eq:7}
  \Delta u = f\text{ on } \Omega, \quad 
n^i\partial_i u  = 0 \text{ on } \partial \Omega,  
\end{equation}
where $n^i$ is the outward unit normal to $\partial \Omega$, the index
$i$ takes values $i=1,\cdots, n$ and $\partial_i$ denotes partial
derivative with respect to the $\Rn$ coordinate $x_i$. 

There exist two main differences between the Neumann and the Dirichlet
problem: (i) The solution to the Neumann problem  is not  unique, for a
given solution we can add a constant and obtain a new solution.
Moreover, the constants are the only solutions of the homogeneous
problem 
\begin{equation}
 \label{eq:6}
  \Delta u = 0 \text{ on } \Omega, \quad 
n^i\partial_i u  = 0 \text{ on } \partial \Omega.  
\end{equation} 
To see this, we multiply \eqref{eq:6} by $u$  and use the divergence
theorem 
\begin{align}
  0= \int_\Omega u \Delta u & = \int_\Omega \partial^i (u \partial_i u
  )-\partial^i u \partial_i u\\
  & = \oint_{\partial \Omega} u n^i\partial_i u - \int_\Omega
  \partial^i u \partial_i u \\ 
  &= -\int_\Omega \partial^i u \partial_i u.
\end{align}
(ii) The source  $f$ can not be arbitrary. We integrate in
$\Omega$  equation \eqref{eq:7} to obtain a necessary condition for
$f$ 
\begin{equation}
\label{eq:9}
0= \oint_{\partial \Omega} n^i\partial_i u =
\int_{\Omega} \Delta u  = \int_\Omega f.   
\end{equation}

The following theorem says  that  \eqref{eq:9} is also a sufficient
condition for the existence of solution.
\begin{theorem}
  A solution $u$  to the Neumann \eqref{eq:7} problem exists if and only if $f$ satisfies
  \begin{equation}
\label{eq:8}
    \int_\Omega f =0.
  \end{equation}
Two different solutions differ by a constant. 
\end{theorem}

The fact that the solution is not unique in the Neumann problem does not  affect the
physics of the model that is described by these equations. Take, for example,
Electrostatics. The electric field $E^i$ satisfies
\begin{equation}
E_i=\partial_i u , \quad \partial_i E^i =f,  
\end{equation}
where $u$ is the electric potential and $f$ the charge. 
If we prescribe $E^in_i$ at the boundary we get a Neumann boundary
problem for the potential $u$. The electric field  $E^i$ is invariant under the transformation
$u\rightarrow u+c$, where $c$ is a constant. We will see in section
\ref{sec:es} that something similar happens for the constraint
equations in  General Relativity.

We have seen that the Neumann problem has not a unique solution. 
If we include lower order terms in the operator, the Dirichlet problem
will not have a unique solution either. For example, 
for some constants $\lambda> 0$ (the eigenvalues) the following equations have
a non-trivial solutions (eigenfunctions)
\begin{equation}
  \Delta u +\lambda u = 0,  \text{ on } \Omega 
\quad u = 0\text{ on } \partial \Omega.
\end{equation} 

One of the main ideas in the theory of partial differential equations is that
many relevant properties of the equations depends only on the principal part,
that is on the terms with highest derivatives. The previous examples show
that uniqueness does not depend only on the principal part. Motivated by the
Neumann problem, we write the following 
the two main  properties of elliptic equations 
\begin{itemize}
\item[(i)] The solutions space of the homogeneous problem (i.e., when we set
the source $f$ and the boundary values $g$ equal to zero) is
  finite dimensional.

\item[(ii)] The solution will exist if and only if the sources satisfy a finite
number of conditions. 
\end{itemize}
We will see in the next sections that, under appropriate assumptions,
(i)-(ii) depend only on the principal part of the equation and
boundary conditions.

One example of a boundary condition that does not satisfy (i) is
the following. 

\begin{example}
\label{ex:tang}
Let $\Omega$ be the
  unit ball in $\Rt$ centered at the origin. 
An explicit calculation shows that the space of solutions of the
homogeneous problem
\begin{equation}
  \Delta u = 0 \text{ on } \Omega, \quad 
\partial_3 u = 0 \text{ on } \partial \Omega, 
\end{equation}
is \emph{infinite dimensional} (see \cite{Popivanov97}, Chapter 1, for details).
Note that the vector $\partial_3$ is tangential to the
boundary at the points $x_3=0$, $x_1^2+x_2^2=1$.
  
\end{example}

\section{Second order elliptic equations}
\label{sec:so}

Consider the following, second order, differential operator 
\begin{equation}
\label{eq:L}
    Lu= \partial_i \left( a^{ij}(x) \partial_j u+ b^i(x) u \right ) +c^i(x)
    \partial_iu + d(x) u,
  \end{equation}
where we will assume that the coefficients are smooth functions on
$\Rn$ and $i,j=1,\cdots n$.  We have
written the operator \eqref{eq:L} in divergence form because it will be more
suitable for the following calculations; since the coefficient $a^{ij}$ and
$b^i$ are 
smooth, this is equivalent to the standard formula 
\begin{equation}
  Lu=  a^{ij}(x) \partial_i \partial_j u+ \hat b^j(x)  \partial_ju + \hat d(x) u. 
\end{equation}
where $\hat b^j= \partial_j a^{ij} + b^j + c^j$ and $\hat d = \partial_i b^i +
d$. 

The principal part of the operator is given by the terms which contains only
second derivatives 
\begin{equation}
  \label{eq:12}
l(x,\partial) = a^{ij}(x) \partial_i \partial_j .  
\end{equation}
To define the symbol  of $L$  we replace in the principal part each derivative by the component of an arbitrary constant vector in $\Rn$
\begin{equation}
    l(x,\xi)=  a^{ij}(x)\xi_i\xi_j, \quad \xi \in \Rn.
\end{equation}
The symbol $l$ of $L$ is a polynomial of order $2$ in the components of $\xi$. 

We make now the crucial assumption on the symbol. We say that the operator $L$
is \emph{elliptic} in $\bar\Omega$ if 
\begin{equation}
\label{eq:10}
    l(x,\xi) \neq 0 \quad \forall x \in \bar\Omega, \, \xi \in 
  \mathbb{R}^n, \xi\neq 0.
\end{equation}

The next important concept is the \emph{formal adjoint} of $L$. The formal
adjoint $L^t$ is defined  by  the relation
\begin{equation}
\label{eq:La}
  \int_\Omega v Lu=  \int_\Omega uL^t v
\end{equation}
for all $u,v$ of \emph{compact support} in $\Omega$.
In this particular case we have
 \begin{equation}
    L^t v= \partial_j \left( a^{ij}(x) \partial_i v- c^j(x) v \right ) -b^i(x)
    \partial_iv +d(x) v
  \end{equation}
Note that $\Delta=\Delta^t$.

We have already seen in the case of the Laplacian that the solutions of the
homogeneous problem play an important role; in general  $L$ and $L^t$
are different operator and then we have two natural null spaces defined as
\begin{align}
\label{eq:null}
\N(L) &=\{u:  Lu =0 \text{ on } \Omega \text{ and } u =0 \text{ on }\partial
\Omega \}\\
\label{eq:nullb}
\N(L^t) & =\{u:  L^t u =0 \text{ on } \Omega \text{ and } u =0 \text{ on }\partial
\Omega \}.
\end{align}
We can now formulate an existence result for the Dirichlet problem which will
essentially ensures that properties (i)-(ii) are satisfied. 

\begin{theorem}
\label{t1}

(i)  Precisely one of the following statements holds:\\
a) For each $f$ there exist a unique solution of the boundary value
problem
\begin{equation}
  \label{eq:a}
  Lu  = f \text{ on } \Omega, \quad 
  u = 0  \text{ on } \partial \Omega,
\end{equation}
or else\\
b) $\N(L)$ is non-trivial.\\ 
ii) Furthermore, should assertion b) hold, the dimension of
 $\N(L)$ is \emph{finite} and equals the
dimension of  $\N(L^t)$.  \\
iii) Finally, the boundary-value problem \eqref{eq:a} has a solution
if and only if 
\begin{equation}
  \int_\Omega fv =0 \quad \text{ for all } v\in \N(L^t). 
\end{equation}
\end{theorem}

We will consider now the analog of the Neumann problem for $L$. If in the
integration by parts given by \eqref{eq:La} we allow functions $u$ and $v$
which are not of compact support, we have to include  the boundary terms; and we
obtain the following relation which is called the 
\emph{Green formula} for the operator $L$
\begin{equation}
\label{eq:11}
 \int_\Omega v L(u)- u L^t(v) = \oint_{\partial \Omega} vB(u)-u B^t(v), 
\end{equation}
where the differential boundary operators are given by
\begin{equation}
\label{eq:b1}
B(u)=n_ja^{ij} \partial_i u + b^in_i u, \quad  B^t(v)=n_ja^{ij} \partial_i v -
c^in_i v.
\end{equation}
We want to solve the following problem
\begin{equation}
  Lu  = f \text{ on } \Omega, \quad 
B(u) = 0  \text{ on } \partial \Omega. 
\end{equation}
As in the case of the Dirichlet problem, we define the null spaces
\begin{align}
\label{eq:nullc}
  \N(L,B) &=\{u: Lu =0 \text{ on } \Omega \text{ and } B(u) =0 \text{
    on }\partial
  \Omega \}\\
\label{eq:nulld}
  \N(L^t,B^t) & =\{u: L^t u =0 \text{ on } \Omega \text{ and } B^t(u)
  =0 \text{ on }\partial \Omega \}.
\end{align}
We have the following existence result, which looks exactly the same as the
previous theorem if we replace the Dirichlet condition by the new boundary
condition.
\begin{theorem}
\label{t2}
(i)  Precisely one of the following statements holds:\\
a) For each $f$ there exist a unique solution of the boundary value
problem
\begin{equation}
  \label{eq:a1}
  Lu  = f \text{ on } \Omega, \quad
  B(u) = 0  \text{ on } \partial \Omega,
\end{equation}
or else\\
b) $\N(L,B)$ is non-trivial.\\ 
ii) Furthermore, should assertion b) hold, the dimension of
 $\N(L,B)$ is finite and equals the
dimension of  $\N(L^t,B^t)$.  \\
iii) Finally, the boundary-value problem \eqref{eq:a1}--\eqref{eq:b1} has a solution
if and only if 
\begin{equation}
  \int_\Omega fv =0 \quad \text{ for all } v\in \N(L^t,B^t). 
\end{equation}

\end{theorem}

We have written the boundary conditions in the form \eqref{eq:b1} in
order to emphasize that they come naturally from the integration by
parts. It is possible to write them in a perhaps more familiar form.
Define the vector $\beta^i$ by
  \begin{equation}
    \beta^i =n_ja^{ij}.
  \end{equation}
  By the elliptic condition \eqref{eq:10} we have $\beta^in_i\neq 0$,
  that is $\beta^i$ \emph{it is never tangential to the boundary}
  (this excludes example \ref{ex:tang}).  In the
  operator $L$ only enters the symmetric part of the matrix $a^{ij}$,
  however, we have not assumed that this matrix is symmetric in the
  previous theorem. If we decompose $a^{ij}=a_s^{ij}+b^{ij}$ where
  $a_s^{ij}=a_s^{(ij)}$ and $b^{ij}=b^{[ij]}$ is an arbitrary anti
  symmetric matrix, then
 \begin{equation}
    \beta^i =n_ja_s^{ij} +\tau^i, \quad \tau^i= n_jb^{ij},
  \end{equation}
where $\tau^i$ is an arbitrary tangential vector.  Choosing
appropriated $b^i$ and $c^i$ such that they do not change the operator $L$, we
get that the function $\sigma= b^in_i$ is also arbitrary. We conclude
that the boundary condition $B(u)=0$ is equivalent to
\begin{equation}
B(u)=\beta^i\partial_iu +\sigma u=0,
\end{equation}
 where $\sigma$ is an arbitrary function and $\beta^i$ is an arbitrary non
tangential vector field on the boundary.

Let us compare theorem \ref{t1} and \ref{t2} with the analog cases
for the Laplace equation. We have now two operators $L$ and $L^t$
which have two different null spaces (in the case when $b^i+c^i=0$ we
have $L=L^t$ and $B=B^t$, and then only one null space).  There are no
statements about uniqueness or about the elements and dimension of the
null spaces. We have already seen that these properties depend on the
lower order terms. For the particular case of second order elliptic operators,
there exist an important tool that can give uniqueness and a
characterization of the null space for certain kind of lower order
terms: the \emph{maximum principle}. There exist many useful versions
of the maximum principle (see for example \cite{Gilbarg}), here we
mention a particular simple case, which can be generalize to other
situations as we will see.

We can write the Green formula \eqref{eq:11} in terms of a first order
bilinear form $\B$
\begin{equation}
 \B(u,v)= \oint_{\partial \Omega} vB(u)- \int_\Omega v L(u) =
 \oint_{\partial \Omega} u B^t(v) -\int_\Omega u L^t(v)
\end{equation}
where
\begin{equation}
  \B(u,v)=\int_\Omega  (a^{ij} \partial_j u + b^i u)\partial_i v - (c^i\partial_i u +
    d u )v .
\end{equation}
From this equation we deduce that $u\in \N(L,B)$ if and only if $\B(u,v)=0$ for all
$v$. (One if is trivial, to see the other one, take test functions $v$ which
vanishes at the boundary and are arbitrary at the interior).
If we assume $b^i=c^i=0$ and $d\leq 0$, then $\B$ is symmetric (i.e. 
  $\B(u,v)= \B(v,u)$) and positive 
\begin{equation}
  \B(u,u)\geq 0, \quad \text{ for all } u.
\end{equation}
Moreover, $ \B(u,u)=0$ if and only if $u$ is a constant and $u=0$ if $d$ is
not identically zero.
In this case we are in a similar situation as in the Neumann problem for the
Laplace equation: the only elements of the null space are the constants.   
 More general version of the maximum principle can be used to prove the
followings refinements of theorems  \ref{t1} and \ref{t2}. 
\begin{theorem}
    Assume $d\leq 0$. Then the Dirichlet problem 
 \begin{equation}
  Lu =f \text{ on } \Omega, \quad 
u =g \text{ on }\partial \Omega,
\end{equation}
has a unique solution for every $f$ and $g$.
\end{theorem}

\begin{theorem}
\label{te:obli}
    Assume $d\leq 0$, $\sigma\geq 0$ and not both identically
    zero. Let $\beta^i$ a vector field such that $\beta^in_i>0$ on
    $\partial \Omega$. 
    Then the oblique derivative problem
 \begin{equation}
  Lu = f \text{ on } \Omega, \quad 
\beta^i\partial_i u +\sigma u =g \text{ on }\partial \Omega,
\end{equation}
has a unique solution for every $f$ and $g$.
\end{theorem}
In both theorems, the maximum principle can be used also to prove that the
solution is positive if the sources, boundary values (and $\sigma$ in theorem
\eqref{te:obli}) are positive.

Note that in theorems \ref{t1} and \ref{t2}  the  null spaces for the operator
and the adjoint have the same dimension, we will see in the next section that
this will not be the case for more general operators and boundary conditions.

We conclude this section with some examples. 

\begin{example}
\label{ex1}
The most important second order elliptic 
operator is the Laplacian on a Riemannian manifold. It is given by
  \begin{equation}
    Lu = \Delta_h u= h^{ij}D_i D_j u,
  \end{equation}
where $h$ is a Riemannian metric ($a^{ij}=h^{ij}$) and $D$ its
corresponding covariant derivative. One important example of lower order term is given by
the conformal Laplacian which appears naturally in the Einstein constraint
equations
\begin{equation}
   Lu = \Delta_h u -\frac{R}{8} u,
\end{equation}
where $R$ is the Ricci scalar of $h_{ab}$.

For a Riemannian metric, the principal part of the boundary condition $B(u)$
has a geometric interpretation
\begin{equation}
 B(u)= n^iD_iu, 
\end{equation}
where we  use the standard convention $n^i=h^{ij}n_j$. That is, the vector $n$
is now the unit normal vector with respect to the metric $h_{ij}$. This is
sometimes denoted as \emph{conormal boundary condition}. 

An example of lower order boundary terms is the following
\begin{equation}
 B(u)= n^iD_iu + Hu, 
\end{equation}
where $H$ is the mean curvature of the boundary $\Omega$ with respect
to the metric $h_{ij}$. This boundary condition appears in connection to
black holes (see \cite{Maxwell03} and \cite{Dain03}).
\end{example}

\section{Elliptic Systems}
\label{sec:es}
\subsection{Definition of ellipticity}
We saw in the previous section that ellipticity is a positivity condition on the
symbol of  the equation. 
In order to generalize this concept  for systems of equations
(this includes as particular case higher order equations) we need to define the
symbol of a system. We can use the same idea as before, and define the
principal part as the collection of terms which have the highest order
derivatives. That is, consider the following differential operator in $\Rn$
\begin{equation}
\label{eq:cop}
L(u)=\sum_{|\alpha|\leq 2m} a_\alpha(x) \partial^\alpha u,
\end{equation}
where $\alpha$ is a multi-index, and the coefficients $a_\alpha$ are $N\times N$
matrices.
The principal part is defined as 
\begin{equation}
\label{eq:cl}
l(x,\partial)=\sum_{|\alpha|= 2m} a_\alpha(x) \partial^\alpha,
\end{equation}
and the symbol
\begin{equation}
l(x,\xi)=\sum_{|\alpha|= 2m} a_\alpha(x) \xi^\alpha.
\end{equation}
The operator is elliptic if $\det l(x,\xi)\neq 0$ for every $x\in
\bar\Omega$ and $\xi\neq 0$. This is the definition that appears in
most text books, we will call it \emph{classical ellipticity} (there
is no general agreement on the nomenclature, in most places these
systems are called just elliptic). This definition excludes many
important examples,  the most
remarkable is perhaps the Laplace equation as a first order system
(example \ref{ex:lfo}).  In order to include these cases, we need to
be more flexible in our definition of the principal part, in particular it
is important to allow terms of different orders in it. This is
particular feature of systems which does not appear in higher order
equations.

It will be convenient to use a more explicit notation as the one given in
\eqref{eq:cl}. 
Let \emph{$u_1,\cdots, u_N$} be  functions which depend on the
coordinates \emph{$x_1,\cdots,x_n$}. The operator \eqref{eq:cop} can be written as
follows
\begin{equation}
\label{eq:14b}
  L_{\mu \nu}(x,\partial)u^\nu(x)=f_\mu(x), \quad \nu,\mu =1,\cdots, N;
\end{equation}
where $L_{\mu\nu}$ are polynomials in $(\partial_1\cdots, \partial_n)$
with
coefficients depending on $x$.
$[L_{\mu\nu}]$ is a $N\times N$ matrix, not necessarily symmetric.
Note that $N$ (dimension of the vectors $u^\nu$) and
$n$ (dimension of $\Rn$) are in general different
numbers.

Let $s_1,\cdots , s_N$, $t_1,\cdots , t_N$ be integers (some may be
negative) such that
\begin{equation}
\label{eq:st}
  \deg (L_{\mu\nu})\leq s_\mu+t_\nu.
\end{equation}
where $\deg$ means the degree of the polynomial $L_{\mu\nu}$ in the
derivatives $\partial$.  The integers $s_\mu$ are attached to the
equations and the $t_\nu$ to the unknowns.

We define the principal part $l_{\mu\nu}(x,\partial)$ as the terms in
$L_{\mu\nu}$ which are \emph{exactly} of order $s_\mu+t_\nu$. The symbol
 $l_{\mu\nu}(x,\xi)$ is obtained replacing in the principal part the derivatives by a
vector $\xi$. We define the following polynomial in
$\xi$
\begin{equation}
\label{eq:det}
  l(x,\xi)=\det(l_{\mu\nu}(x,\xi)). 
\end{equation}
The degree $m$ of the systems is given by 
\begin{equation}
\label{eq:13b}
  m=\frac{1}{2} \deg(l(x,\xi)),
\end{equation}
where $\deg$ means degree in $\xi$. 
 
The following general definition of ellipticity was introduced in \cite{Douglis55}
\begin{definition}[Douglis-Nirenberg Ellipticity]
\label{dn}
The system \eqref{eq:14b} is elliptic if there exist integer weights $s_\mu$ and
$t_\nu$ which satisfy  \eqref{eq:st} and such  $l(x,\xi)\neq 0$ for all real $\xi\in\Rn$,
 $\xi\neq 0$, $x\in \bar \Omega$; where  $l(x,\xi)$ is given by \eqref{eq:det}.  
\end{definition}

For $n=2$ we assume in addition
\begin{definition}[Supplementary condition]
\label{sc}
$l(x,\xi)$ is of even degree $2m$. For every pair of linearly independent
real vectors $\xi$ and $\xi'$, the polynomial $l(x,\xi+\tau \xi')$ in
the complex variable $\tau$ has exactly $m$ roots with positive
imaginary parts.   
\end{definition}

Every elliptic systems in dimension $n\geq 3$ satisfies the
supplementary condition (see \cite{Agmon59}). This no longer true for $n=2$, as example \ref{ex:cr}
shows. A system that is elliptic in the sense of Definition \ref{dn} and
satisfies also the supplementary condition (Definition \ref{sc}) will be called
\emph{properly elliptic}. 

Note that the definition depends on the weights $s_\mu$ and $t_\nu$ which
are not unique, a system can be elliptic for many different choices of
weights. Also note that the number $2m$ is not related in general
with the degree of the highest derivatives, for example for a second
order system with $N=3$ we have $m=3$ (example \ref{ex:york}). The
degree $m$ is important because it gives the number of boundary
conditions we have to impose in order to get a well defined elliptic
problem, as we will see in the next section.

There exists an  important class of elliptic operators for which
the Dirichlet boundary conditions will always satisfy (i)-(ii) as we
will see in the next section. These systems are given by the following
definition. 

\begin{definition}[Strong Ellipticity]
The system is called strongly elliptic if $s_\nu=t_\nu \geq 0$ and
there exist a constant $\epsilon >0$ such that
\begin{equation}
 Re \, (l_{\mu\nu}(x,\xi)\eta^\mu \bar \eta^\nu) \geq \epsilon \eta^\mu\eta_\mu \xi^i\xi_i, 
\end{equation}
  for all real $\xi\in\Rn$ and all complex $\eta\in \mathbb{R}^N$.
\end{definition}
Note that every elliptic equation (i.e., $N=1$) is strongly elliptic. 
Let us discuss some examples. 

\begin{example}[Laplace equation as a first order system]
\label{ex:lfo}
This example was taken from \cite{Agmon64}. 
Consider the Laplace equation in two dimensions
  \begin{equation}
  \partial_1^2u +\partial_2^2u=0.
   \end{equation}
Every equation can be written as a first order system if we introduce the
derivatives of the unknown as new variables. That is, let $u_1=\partial_1u$
and  $u_2=\partial_2 u$. Then we have the following system ($n=2$ and $N=3$)
\begin{align}
     \partial_1 u_1+ \partial_2 u_2 &=0,\\
     \partial_1 u -u_1 &=0,\\
     \partial_2 u -u_2 &=0.
 \end{align}
In the matrix notation
\begin{equation}
  \begin{pmatrix} 
    0 & \partial_1 & \partial_2 \\
    \partial_1 & -1 & 0 \\
\partial_2 & 0 & -1
  \end{pmatrix} 
  \begin{pmatrix} 
    u \\
    u_1 \\
u_2\\
  \end{pmatrix}  =0.
\end{equation}
In the  classical definition, the symbol is constructed only with the terms
which contains the highest order derivatives, in this case only with the terms
with one derivative. Then the determinant of the symbol is
\begin{equation}
  \begin{vmatrix} 
    0 & \xi_1 & \xi_2 \\
    \xi_1 & 0 & 0 \\
\xi_2 & 0 & 0
  \end{vmatrix} =0,
\end{equation}
and we conclude that the  system is not classically elliptic. 

Take the weights   $t_1=2$,  $t_2=t_3=1$, for $u,u_1,u_2$ and
$s_1=0$, $s_2=s_3=-1$, 
to the first, second and third equations, respectively.
Then, we have
\begin{equation}
  \begin{vmatrix} 
    0 & \xi_1 & \xi_2 \\
    \xi_1 & -1 & 0 \\
\xi_2 & 0 & -1
  \end{vmatrix} =\xi_1^2+\xi_2^2,
\end{equation}
and the system is elliptic with $m=1$. Another possible choice for the
weights is the following $s_i=t_i$, with $t_1=1$, $t_2=t_3=0$.

Since $n=2$, we have to check also that it satisfies the supplementary condition
\begin{equation}
 0= l(\xi+\tau\xi')=|\xi|^2+2\tau\xi^i\xi'_i+\tau^2|\xi'|^2
\end{equation}
then
\begin{equation}
  \tau_{\pm}=(-\cos\theta \pm i\sin\theta)|\xi'|^{-1}|\xi|
\end{equation}
where $|\xi|^2=\xi^i\xi_i$ and $\xi^i\xi'_i=\cos\theta |\xi||\xi'|$.
That is, we have only one root with positive imaginary part.

\end{example}

\begin{example}[Stokes system]
\label{ex:stokes}
This example was taken from \cite{Renardy04}.
The following equations appear as the stationary linearized case of
the Navier-Stokes equations (see for example \cite{Sohr01}) for the
velocity $u^i$ and the pressure $p$ of the fluid
\begin{equation}
\label{eq:stokes}
    \Delta u^i-\partial^i p =0, \quad \partial^iu_i =0.
  \end{equation}
The unknowns are $u^i,p$, that is $N=4$, and we will assume
$n=3$. Then, in the matrix notation we have 
\begin{equation}
\begin{pmatrix} 
\Delta    & 0  & 0 & -\partial_1  \\ 
   0   & \Delta  & 0 & -\partial_2  \\
0    & 0  & \Delta & -\partial_3  \\
\partial_1    & \partial_2  & \partial_3 & 0  \\
  \end{pmatrix} 
  \begin{pmatrix} 
    u_1 \\
    u_2 \\
u_3\\
p \\
  \end{pmatrix}=0.
\end{equation}
It is clear that the system is not classically elliptic. 
Take $t_1=t_2=t_3=2$, $t_4=1$ and  $s_1=s_2=s_3=0$,
$s_4=-1$. Then the symbol is
\begin{equation}
l_{ij}=\begin{pmatrix} 
|\xi|^2    & 0  & 0 & -\xi_1  \\ 
   0   & |\xi|^2  & 0 & -\xi_2  \\
0    & 0  & |\xi|^2 & -\xi_3  \\
\xi_1    & \xi_2  & \xi_3 & 0  \\
  \end{pmatrix}, 
\end{equation}
and we have
\begin{equation}
  l = |\xi|^6, \quad  m =3.
\end{equation}
Then, the system is elliptic.  
Another possible choice for the weights is the following:
$s_i=t_i$, with $t_1=t_2=t_3=1$ and $t_4=0$.

\end{example}

\begin{example}[Cauchy-Riemann equation]
\label{ex:cr}
We write the Cauchy-Riemann equation 
$L(u)=\partial_{\bar z} u$, in terms of the real variables $z=x+iy$
\begin{equation}
  L(u)=\frac{1}{2} \left( \partial_x u + i\partial_y u\right).
\end{equation}
We have $n=2$, $N=1$. The symbol  $l=\xi_1+i\xi_2$, satisfies 
\begin{equation}
  l(\xi)\neq 0 \text{ for all real }\xi\neq 0,
\end{equation}
hence the system is elliptic with $m=1/2$. 
However, it does not satisfies the supplementary condition because
$2m=1$ is not an even number.
\end{example}

\begin{example}
\label{ex:york}
Consider the following operator in $\Rt$, acting on three vectors $u^i$
\begin{equation}
\label{eq:lo}
L_{ij}u^j=\partial^j(\ek  u)_{ij},
\end{equation}
where 
\begin{equation}
\label{eq:cko}
  (\ek  u)_{ij}=2\mu\partial_{(i}u_{j)}+\lambda\delta_{ij}\partial^ku_k,
\end{equation}
and $\mu,\lambda$ are constants. Since in this case we have $N=n=3$ we
will use the same index notation for the index in the vectors $u$ and
in the coordinates of $\Rt$.

The system \eqref{eq:lo} appears in
elasticity (see, for example, \cite{Marsden83}). It also appears in
General Relativity related to gauge conditions like the minimal
distortion gauge (see \cite{Smarr78a}) and in the constraint equations
(see \cite{York73}), usually with the choice $\mu=1$, $\lambda=-2/3$
which makes \eqref{eq:cko} trace free.

From \eqref{eq:lo} we deduce
\begin{equation}
   L_{ij}u^j = \left((\mu+\lambda)\partial_i\partial_j
     +\mu\delta_{ij}\Delta\right)u^j, 
\end{equation}
in the matrix notation  we have ($\lambda'=\mu+\lambda$)
\begin{equation}
  L_{ij}u^j\equiv  \begin{pmatrix} 
    \lambda'\partial^2_1 +\mu\Delta    &\lambda' \partial_2
    \partial_1 & \lambda'\partial_1  \partial_3 \\ 
    \lambda' \partial_1  \partial_3 & \lambda'\partial^2_2
    +\mu\Delta & \lambda' \partial_1  \partial_3  \\
    \lambda' \partial_1  \partial_3 & \lambda' \partial_1
    \partial_3 & \lambda'\partial^2_3 +\mu\Delta 
  \end{pmatrix} 
  \begin{pmatrix} 
    u_1 \\
    u_2 \\
u_3\\
  \end{pmatrix}.
\end{equation}
Take $s_i=t_i=1$, the symbol is given by
\begin{equation}
  l_{ij}(\xi)=\lambda'\xi_i\xi_j +\mu\delta_{ij} \xi^k\xi_k, 
\end{equation}
and 
\begin{equation}
  l=\mu^2(2\mu+\lambda)|\xi|^6, \quad m=3. 
\end{equation}
The operator is (classically) elliptic for $\mu> 0$,
$2\mu+\lambda > 0$. 
It is also strongly elliptic
\begin{equation}
   l_{ij}\eta^i\bar \eta^j=(\lambda+\mu) (\eta^i\xi_i)(\bar \eta^i\xi_i)+\mu
   \xi^k\xi_k\eta_i\bar \eta^i\geq \epsilon \xi^k\xi_k \eta_i\bar \eta^i, 
\end{equation}
where $\epsilon=\min\{\mu, 2\mu + \lambda\}$.
\end{example}

\begin{example}[Einstein Constraint equations]
  
  There exist different ways of reducing the Einstein constraint
  equations to an elliptic systems (see, for example, the recent review
  \cite{Bartnik:2002cw}).  In the standard approach the principal part
  of the system is formed with the Laplace operator on a Riemannian
  manifold given in example \ref{ex1} and the operator that has been
  discussed in example \ref{ex:york}.

  A particular interesting example is the one that has been recently
  used in \cite{Chrusciel:2003sr} and \cite{Corvino:2003sp} to
  construct new kind of solutions.  This system is not elliptic in the
  classical sense but it satisfies definition \ref{dn} for
  appropriate weights (see these references for details).
 
\end{example}

\begin{example}[Witten equation]
\label{ex:witten}
The Witten equation $\partial_{AA'}u^{A}=0$ (in the spinorial
notation) plays an important role in the positive mass theorem of
General Relativity (cf. \cite{Witten81}). Solutions of this equation has been analyzed in  
\cite{MR83e:83036} and \cite{Parker82}. 

In the matrix notation ($N=2$, and we will assume $n=3$) this system is given by 
\begin{equation}
\label{eq:witten}
  \begin{pmatrix} 
    \partial_3 & \quad \partial_1 + i\partial_2 \\
    \partial_1 -i\partial_2 &\quad  - \partial_3  
  \end{pmatrix} 
  \begin{pmatrix} 
    u_1 \\
    u_2 
  \end{pmatrix}  =0.
\end{equation}
 The principal part, with weights $t_1=t_2=1$, $s_1=s_2=0$, is given by
\begin{equation}
l_{\nu\mu}(x,\xi)=
\begin{pmatrix} 
\xi_3  &  \xi_1+i\xi_2 \\
  \xi_1-i\xi_2 & -\xi_3  \\
\end{pmatrix}, \quad l= -(\xi^2_3+ \xi^2_2+\xi^2_1), \quad m=1.  
\end{equation}
Then, the system is elliptic. 
 
\end{example}

\subsection{Definition of elliptic boundary conditions}

For the operator $L_{\mu\nu}$ defined 
in  \eqref{eq:14b} we will consider boundary conditions of the
following form   
\begin{equation}
B(x,\partial)_{l\nu}u^\nu=0, \quad l=1, \cdots , m; \quad \nu=1, \cdots , N;
 \end{equation}
where $B(x,\partial)_{l\nu}$ are  polynomial in $\partial$ and $m$ is given by 
\eqref{eq:13b}. 
The order of the boundary operators, like those of the operators in \eqref{eq:14b},
depends on two systems of integer weights, in this case the system $t_\nu$
already attached to the dependent variable and a new system 
  $r_l$  attached to each boundary condition such that
\begin{equation}
\deg  (B_{l\nu})\leq r_l+t_\nu. 
\end{equation}
Note that $r_l$ can be negative and also the order of the derivatives in the
boundary conditions can be higher than in the operator. 
The principal part $b_{l\nu}$  of the boundary operator consists of the terms  
in $B_{l\nu}$  which are exactly of  order $r_l+t_\nu$.

For a given operator $L$, we would like to know for which boundary
operators $B$ the solutions of the corresponding boundary value
problem will satisfies (i)-(ii). The answer to this question is given
by the following definition, as we will see in the next section. 

Let $x_0$ a point on $\partial \Omega$ and let $n^i$ the outer normal to
$\Omega$. We consider the constant coefficient problem
\begin{align}
  l_{\mu\nu}(x_0,\partial)u^\nu & =0,\label{eq:ls1}\\
b_{l\nu}(x_0,\partial) u^\nu &=0\label{eq:ls2},
\end{align}
on the half plane $(x^i-x^i_0)\cdot n_i < 0$ with boundary $(x^i-x^i_0) n_i =
0$. 
\begin{definition}[Complementing condition]
\label{cc}
  We say that the complementing condition holds at $x_0$ if there are
  no nontrivial solutions of \eqref{eq:ls1}--~\eqref{eq:ls2} of the
  following form:
  \begin{equation}
\label{eq:exp}
    u^\nu(x)=v^\nu(\eta) e^{i\xi_j (x^j-x^j_0)} 
  \end{equation}
where $\xi$ is a any nonzero, real, vector  which satisfies
$\xi^in_i=0$,  $v(\eta)$ tends to zero exponentially as
$\eta \rightarrow -\infty$ and the coordinate $\eta$ is defined by 
$\eta=(x^j-x^j_0) n_j$. 
\end{definition}
In the literature, these conditions are also called
\emph{Lopatinski-Shapiro} conditions or \emph{covering} conditions (see
\cite{Agmon64} and \cite{Wloka95}).
Let us study some examples of boundary conditions.

\begin{example}[Boundary conditions for the Laplace equation.] 
Consider solutions of the form \eqref{eq:exp} for the Laplace equation $\Delta
u=0$. We chose coordinates in $\Rn$ such that $\eta=x_n$, $n^i=\delta^i_n$.
Then, all the solutions of this form are given by
\begin{equation}
    \label{eq:1}
    u=e^{i\xi^i x_i}e^{\pm |\xi|x_n},
  \end{equation}
where $\xi$ satisfies $\xi_n=0$. 

We consider different boundary conditions on the plane $x_n=0$. For the 
 Dirichlet condition $u(x_n=0)=0$  we get 
\begin{equation}
  \label{eq:2}
   u=e^{i\xi^ix_i}=0,
\end{equation}
since this is not possible there exist no solution of this form  which satisfies
the Dirichlet conditions. Hence, the Dirichlet boundary condition satisfies the
complementing condition. 

For the Neumann condition we have $\partial_{x_n} u = 0$ at  $x_n=0$, this implies $\xi=0$ and then the
solution will not decay at infinity. Hence, the Neumann conditions satisfied
also the complementing conditions. 

Take the oblique derivative boundary condition  $\beta^i\partial_i u =0$ at
$x_n=0$. This implies
\begin{equation}
  \label{eq:4}
  i(\beta_i\xi^i)=0, \quad \beta_n |\xi|=0.
\end{equation}
If $\beta_n\neq0 $ then $|\xi|=0$, an the complementing condition is
satisfied.  This was the case studied in section \ref{sec:so}.  On the
other hand, if $\beta_n=0$ (like in example \ref{ex:tang}) then the
complementing condition is not satisfied since we can always chose a
vector $\xi$ such that $\beta_i\xi^i=0$ and we will get solutions of
the form \eqref{eq:exp}.

Consider now the following  interesting example studied in \cite{Hoermander63}. 
At $x_n=0$ we impose the boundary conditions
\begin{equation}
  \label{eq:5}
  \delta u =0,
\end{equation}
where
\begin{equation}
  \label{eq:5b}
  \delta u = \partial^2_1u + \cdots + \partial^2_{n-1}u,
\end{equation}
is the Laplacian in $n-1$ dimension. From \eqref{eq:5} we deduce the $|\xi|^2=0$
and then it satisfies the complementing conditions. It is also clear that
$\delta^k u =0$ where $k$, is an arbitrary natural number, satisfies the
complementing condition. Note that in this cases the boundary operator has
derivatives of higher order than the Laplace operator. On a Riemannian manifold,  these conditions can be written in
geometric form where $\delta$ is the intrinsic Laplacian on the boundary.
Another  interesting
condition which also satisfies the complementing condition is the following
 \begin{equation}
  \label{eq:5h}
  \delta u-n^i \partial_i u =0.
\end{equation} 
In this case, integrating by parts, it  is easy to show that the only solutions of the homogeneous problem
are the constants
  \begin{align}
  \label{eq:5c}
0=\int_{\Omega} u \Delta u &= \oint_{\partial\Omega}un^i\partial_i u -\int_\Omega
\partial_i u \partial^i u\\
& = \oint_{\partial\Omega}u \delta u -\int_\Omega
\partial_i u \partial^i u\\
& = - \oint_{\partial\Omega}| d u|^2 -\int_\Omega
\partial_i u \partial^i u,
\end{align}
where $du$ denotes the gradient intrinsic to the boundary.

\end{example}

\begin{example}
Consider the operator discussed in example \ref{ex:york}. Integrating
by parts we get 
\begin{equation}
\label{eq:bp}
  \B(u,v)= -\int_\Omega v^iL_{ij}u^j+ \oint_{\partial \Omega}
 (\ek u)_{ij}n^i v^j
\end{equation}
where
\begin{equation}
  \B(u,v)= \int_\Omega  (\ek u)^{ij}  \partial_i v_j. 
\end{equation}
We can write the integrand in $\B(u,v)$ in the following form
\begin{equation}
\label{eq:int}
 (\ek u)^{ij}  \partial_i v_j= \frac{\mu}{2} (\ck u)_{ij} (\ck v)^{ij}
 +(\lambda + \frac{2}{3}\mu) \partial_k u^k \partial_l v^l, 
  \end{equation}
where $(\ck u)_{ij}$ is the trace free part of $\partial_{(i} u_{j)}$,
that is 
\begin{equation}
  \label{eq:14}
  (\ck u)_{ij}= 2 \partial_{(i} u_{j)}-\frac{2}{3}\delta_{ij} \partial_k u^k.
\end{equation}
Note that $\B$ is symmetric $\B(u,v)=\B(v,u)$.  
Using this and equation \eqref{eq:bp} we get 
the following Green  formula
\begin{equation}
\label{eq:green2}
 \int_\Omega v^iL_{ij}u^j-  u^iL_{ij}v^j=  \oint_{\partial \Omega}
 (\ek u)_{ij}n^i v^j- (\ek v)_{ij}n^i u^j.
\end{equation}
This is analogous to the Green formula for second order equations
 \eqref{eq:11}. For
simplicity we have not included terms in non divergence form in the operator,
that is why we have $L=L^t$ and $B=B^t$ in \eqref{eq:green2}, these extra terms
can be handle in the same way as in section \ref{sec:so}. 

The boundary integral in the Green formula \eqref{eq:green2} suggests that two natural boundary
conditions are the Dirichlet 
\begin{equation}
\label{eq:diriy}
u^i=0\text{ on } \partial \Omega,
\end{equation}
and the analog to the Neumann boundary
condition
\begin{equation}
\label{eq:neu}
  (\ek  u)_{ij}n^j = 0 \text{ on } \partial \Omega.
\end{equation}
We want to prove that these boundary conditions satisfy the complementing
conditions. We will assume that $\mu>0$ and $2\mu+\lambda>0$, that is, the
operator is elliptic as we have seen in example  \ref{ex:york}. We will make
also an extra assumption:  $3\lambda + 2\mu\geq 0$; this implies that the
integrand  \eqref{eq:int}
is positive. Moreover, if  $3\lambda + 2\mu> 0$
\begin{equation}
B(u,u)=0 \iff \partial_{(i}u_{j)}=0,
\end{equation} 
that is $u$ is a Killing vector. If $3\lambda + 2\mu= 0$, then
\begin{equation}
B(u,u)=0 \iff (\ck u)_{ij}=0,
\end{equation}
then $u$ is a conformal Killing vector. In flat space, we know
explicitly all the Killing and conformal Killing vectors. Hence, we
have a characterization of the null spaces for these boundary
conditions. The Killing and conformal Killing are the analog of the
constants for the Neumann problem for the Laplace equation.

Assume we have a solution $u$ of the form \eqref{eq:exp}.  Chose
Cartesian coordinates such that $\eta=x_3$. Let $L_1=2\pi/\xi_1$ and
$L_2=2\pi/\xi_2$. Take as domain the infinite cubic region $x_3\geq 0$,
$0\leq x_1 \leq L_1$, $0\leq x_2 \leq L_2$.  For this domain we use
equation \eqref{eq:bp} for $u=v$. We want to prove that, on
this domain, the boundary integral in \eqref{eq:bp} vanished if we
impose either \eqref{eq:diriy} or \eqref{eq:neu}. Using these boundary
conditions we get that the integrand vanishes on the face $x_3=0$. The
integrand also vanishes on the face $x_3=\infty$ because the solution,
by hypothesis, decay at infinity. On the other faces the integrand
does not vanish. However, because of the choice of $L_1$ and $L_2$, we
have that the integrand of opposite faces are identical. Then, the
sum of the boundary integrals vanished because the normal is always
outwards.  We conclude that  $\B(u,u)$ should vanish. But
there are no Killing or conformal Killing vectors which decay to zero
at infinity. Hence the complementing condition is satisfied.
 
\end{example}
\begin{example}[Boundary conditions for the Stoke system]
If we multiply equations \ref{eq:stokes}  by $u^i$ and integrate by parts we get
\begin{equation}
0=-\int_\Omega\partial_k u_i \partial^k u^i + \oint n^k (u^i\partial_k u_i -
u_k p).
\end{equation}
Using this equation and a similar argument as in the previous example it is
possible to show that the boundary conditions
\begin{equation}
u^i=0 \text{ on } \partial \Omega,
\end{equation}
and $p$ unprescribed, satisfy the complementing condition (see for example
\cite{Renardy04}).
\end{example}

\begin{example}[Dirichlet boundary conditions for strongly elliptic
  systems]

Assume that the system is strongly elliptic, this implies
$s_i=t_i=t'_i\geq 0$. The Dirichlet boundary conditions on $\partial
\Omega$ are given by
   \begin{equation}
\label{eq:diri}
   (n^i\partial_i)^qu_j=0, \quad q=0,\cdots,
   t'_j-1,\quad  j=1,\cdots , N; 
  \end{equation}
when $t'_j=0$, $u_j$ goes unprescribed.

It can be proved that for every strongly elliptic system, the Dirichlet
conditions \eqref{eq:diri} satisfy the complementing condition (see
\cite{Agmon64}).
 
In the case equations ($N=1$) of order $2m$ ($t_1=m$) these conditions reduce to 
   \begin{equation}
\label{eq:diri2}
   (n^i\partial_i)^qu=0, \quad q=0,\cdots, m-1.
  \end{equation}
In particular, for second order equation ($m=1$) we have $u=0$ at the boundary.
That is, we recover the familiar Dirichlet condition studied in sections
\ref{sec:in}
and \ref{sec:so}. 
In example \ref{ex:york} we have
 $t'_i=1$, then $q=0$ and the Dirichlet conditions is just
\begin{equation}
  u^j=0 \text{ on } \partial \Omega.
\end{equation}

As an example of a higher order equation, we have the
biharmonic equation
  \begin{equation}
\label{eq:bh}
    \Delta \Delta u=f,
  \end{equation}
the Dirichlet conditions are  given by  ($N=1$, $m=2$) 
\begin{equation}
  u=0, \quad n^i \partial_i u =0 \text{ on } \partial  \Omega.
\end{equation}

\end{example}

\begin{example}
In the following example (taken from \cite{Renardy04}), the complementing condition is
\emph{not} satisfied.
  Consider the following problem in $\mathbb{R}^2$, where the boundary
  is the line $x_2=0$
  \begin{equation}
    \Delta \Delta u =0 \text{ on } \Omega \quad
\Delta u = \partial_2 \Delta u=0 \text{ on } \partial \Omega.
  \end{equation}
For every $\xi \in \mathbb{R}$ the function
\begin{equation}
u(x,y)=e^{i\xi x_1-|\xi|x_2}
\end{equation}
is a solution.
  
\end{example}

\begin{example}
We have seen that for strongly elliptic system the Dirichlet boundary
conditions  satisfy the complementing condition. This is not true for general
elliptic systems. In  following example  (discussed in \cite{Nirenberg56}) we
show that there are elliptic systems for which the Dirichlet problem is not
well defined. 

Consider the system ($N=n=2$) 
\begin{equation}
  \begin{pmatrix} 
   \partial_1^2 - \partial_2^2 &\quad  -2\partial_1\partial_2 \\
 2\partial_1\partial_2 & \quad \partial_1^2 - \partial_2^2  
  \end{pmatrix} 
  \begin{pmatrix} 
   u_1 \\
u_2\\
  \end{pmatrix}  =0.
\end{equation}
The symbol is given by
\begin{equation}
 l_{ij}=\begin{pmatrix} 
   \xi_1^2 - \xi_2^2 &\quad  -2\xi_1\xi_2 \\
 2\xi_1\xi_2 & \quad \xi_1^2 - \xi_2^2  
  \end{pmatrix}, \quad  l = (\xi_1^2 + \xi_2^2)^2. 
\end{equation}
Then, the system is elliptic in the classical sense. This system  can
be written in the complex form, $z=x_1+ix_2$, $w=u_1+iu_2$, as 
\begin{equation}
\partial_{\bar z}^2 w =\frac{1}{4} (\partial_{1}+ i\partial_{2})^2 w=0,
\end{equation}
for which the general solution is clearly
\begin{equation}
w=f(z)+\bar z g(z),
\end{equation}
where $f$ and $g$ are arbitrary functions of $z$.
We observe that all solutions of the form 
\begin{equation}
w=f(z)(1-z\bar z) \text{ on } |z|\leq 1
\end{equation}
with arbitrary analytic $f$, vanish on the boundary of the unit disk. Thus the
problem of finding a solution with Dirichlet boundary condition is not well
defined. 
\end{example}
\begin{example}[Boundary conditions for the Witten equation]
In the Witten equation studied 
in example \ref{ex:witten} we have $m=1$, that is, we can only impose one boundary
condition. Consider the following boundary condition
\begin{equation}
u_1=0, \text{ on } \partial \Omega,
\end{equation}
and $u_2$ goes unprescribed. This condition has been studied in \cite{reula:84} as an
inner boundary condition for black holes in the positive mass theorem. 
We want to prove that it satisfies the complementing condition. 
We can explicitly calculate all the  solutions of the form
\eqref{eq:exp} of the equations \eqref{eq:witten} 
\begin{equation}
u^\nu=e^{\xi^ix_i}v^\nu(x_3),\quad  v^\nu=A^\nu e^{ |\xi|x_3},
\end{equation}
where $A^\nu$ are constants such that
$A_2/A_1=(i\xi_1+\xi_2)|\xi|^{-1}$ and we chose coordinates such that
$\eta=x_3$, $\xi_3=0$.  There is no solution of this form that satisfies
$u_1(x_3=0)=0$ and then the complementing condition follows.
\end{example}

\begin{example}[Stationary solutions of Einstein equations]
  
  In the presence of a timelike symmetry, Einstein equations can be
  reduced to an elliptic system. Moreover, the inner boundary
  conditions satisfy the complementing condition. This result was
  proved in \cite{Reula89} and it was used to prove an existence
  result for the non linear problem. See also \cite{Miao03} for a
  different kind of boundary conditions for the static case.

\end{example}

\subsection{Results}
In order to present a general result for properly elliptic systems with
boundary conditions that satisfy the complementing condition we need
to reformulate in a more precise way properties (i)-(ii). For a given
operator $L$ and boundary operator $B$ we consider the operator $A$
defined as $A(u)=( L(u),B(u))$. This operator will act on appropriate
Hilbert spaces $\spa{H}_1$ and $\spa{H}_2$, $A:\spa{H}_1\to
\spa{H}_2$.  In analog way as we did in \eqref{eq:nullc} we define the
null space $\N(A)$ as
\begin{equation}
  \N(A) =\{u\in \spa{H}_1 :  A(u) =0  \}.
 \end{equation}
The range of $A$ is defined by
\begin{equation}
  \Ra(A) =\{w\in \spa{H}_2 : \exists  u\in \spa{H}_1,  A(u) =w  \},
 \end{equation}
and the complement of the range is given by
\begin{equation}
  \Rc(A) =\{w\in \spa{H}_2 :  (Au,w)_{\spa{H}_2}=0 \text{ for all }u\in \spa{H}_1  \}.
 \end{equation}
where $(\cdot , \cdot)$ denotes the Hilbert scalar product. 

We can write now properties (i)-(ii) as follows

\begin{itemize}
\item[(i)]  $\N(A)$ has finite dimension.

\item[(ii)] $\Rc(A)$ has finite dimension. 
\end{itemize}
An operator which satisfies (i)-(ii) is called a \emph{Fredholm}
operator.  (we have assumed that $A$ is bounded, otherwise in (ii) we
need to impose that $\Ra(A)$ is closed, see \cite{Kato80} and
\cite{Hoermander85}).

We have the following general result (we only sketch the statement, for details and
proofs see \cite{Hoermander85b},\cite{Hoermander85} and also \cite{Wloka95})
\begin{theorem}
\label{Horm}
If the system $L$ is properly elliptic in $\bar \Omega$ and the
boundary conditions satisfy the complementing condition for every
point of $\partial \Omega$, then the operator $A(u)=(L(u),B(u))$ is
Fredholm.
\end{theorem}

We have seen that the dimension of $\N$ (and hence uniqueness) is  not invariant if we add lower order
terms to the operator. One of the consequence of theorem \ref{Horm} is the
existence of an invariant for elliptic problems: the Fredholm index. This number
is defined as
\begin{equation}
I= \dim\N(A) - \dim\Rc(A).
\end{equation} 
It can be proved that the index $I$ is stable under perturbation, in particular
it does not depend on the lower order terms. 

In section \ref{sec:so} we have used the Green formula to construct
the formal adjoint operator $L^t$ and its corresponding boundary
operator $B^t$. In this case we can define $A^t(u)=(L^t(u),B^t(u))$,
and it can be proved \footnote{It is important to note that for any
  bounded (or unbounded with dense range) operator $A$ we can define
  the Hilbert adjoint $A'$. This is not related, in general, with the
  formal adjoint $A^t$. However, when we have a Green formula, it is
  possible to prove that in fact $A^t=A'$ (see theorem 8.4 of
  \cite{Lions72}, \cite{Browder59} and also \cite{Taylor96}} that
$N(A^t)= \Rc(A)$.  That is, the boundary value problems considered in
theorem \ref{t1} and \ref{t2} have $I=0$. In fact these theorems also
show that the index does not depend on the lower order terms in this
particular case.

Boundary conditions which come  from a Green formula are
called \emph{normal boundary conditions}. The advantage of them is that we have
a characterization of $\Rc(A)$ through the formal adjoint problem, and then we
can in principle compute the conditions that the sources should satisfy in
order to have a solution.
General results for normal boundary conditions for higher order elliptic
equations can be found in \cite{Lions72} \cite{Schechter59}.
 Since these boundary conditions come from an
integration by parts, the order the boundary operators will be always less than
the operator itself. We have seen that this is not necessary the case for
general elliptic boundary conditions that satisfy the complementing condition. 
For the general case, we will not have a characterization of $\Rc(A)$.

We have seen that the Dirichlet boundary conditions satisfy the complementing
condition for strongly elliptic systems. Using this fact, general existence
results for the Dirichlet problem can be proved (see \cite{Nirenberg55}).
Moreover, it can be shown that the index is always zero in this case.

Finally, we want to present an existence result for the operator considered in 
 example \ref{ex:york} that can be deduced from the general theorem \ref{Horm}
(see \cite{Thompson69}) In this case, we have a Green formula and then we have
normal boundary conditions. The following two theorems are the analogous of
theorem \ref{t1} and \ref{t2}. 

\begin{theorem}
  Let $L_{ij}$ be given by \eqref{eq:lo} with $\mu\geq 0$, $2\mu+
  \lambda\geq 0$, $2\mu+3\lambda \geq 0$. 
  Then, for every smooth $f^j$ and $g^i$, there exist a unique,
  smooth, solution $u^i$ of the Dirichlet problem
\begin{equation}
  \label{eq:13c}
  L_{ij}u^i=f^j \text{ on } \Omega, \quad u^i=g^i \text{ on }\partial \Omega.
\end{equation}
\end{theorem}

We have seen that all solution of the homogeneous problem satisfy
$(\ek v)_{ij}=0$, that is $v$ is a Killing or a conformal Killing
vector.  Uniqueness in this theorem follows because there exists no
Killing or conformal Killing vector which vanishes at the boundary.

\begin{theorem}
\label{t4}
 Let $L_{ij}$ be given by \eqref{eq:lo} with $\mu\geq 0$, $2\mu+
  \lambda\geq 0$, $2\mu+3\lambda \geq 0$. Consider the boundary value problem
\begin{equation}
  \label{eq:13d}
  L_{ij}u^i=f^j \text{ on } \Omega, \quad (\ek u)_{ij}u^i =0 \text{ on }\partial \Omega.
\end{equation}
This problem has a solution if and only if
 \begin{equation}
  \int_\Omega f_i v^i =0 \quad \text{ for all } v^i \text{ such that } \ek{v}_{ij}=0. 
\end{equation}
If $u_1$ and $u_2$ are two different solutions, then the difference $v=u_1-u_2$
satisfies $(\ek v)_{ij}=0$.
\end{theorem}
In the case of the Einstein  constraint equations, the previous theorems can be
used to prove existence of solutions of the momentum constraint (see \cite{York73}).
In this case the physical quantity is the second fundamental form $K_{ij}$
which is given by
\begin{equation}
  K_{ij}= Q_{ij}- (\ek u)_{ij},
\end{equation}
where $Q$ is an (essentially) arbitrary tensor. Then, as in the case of the
Neumann problem for the Laplace equation, the lack of uniqueness in theorem
\ref{t4} will not affect $K_{ij}$.

\section{Final Comments}

In order to check if a system of equations is elliptic, we should first prove
that the principal part of the operator satisfy definitions  \ref{dn} and \ref{sc}.
If the system is non linear, we should consider the corresponding linearized
problem. Then we should prove that the boundary operators satisfy the
complementing condition  (definition \ref{cc}). This can be complicated. There
exist equivalent formulations of this condition (see for example \cite{Wloka95}),
some of them can be more suitable for specific problems. It is also important
to know if the boundary conditions come from a Green formula (normal boundary
conditions). The Green formula can be used to prove that the complementing
condition holds (as we have seen in the examples). Moreover, for the case of
higher order equations  there exist general results that can
be used (see \cite{Lions72}). 

We have only discussed linear elliptic systems. In the non linear
case, there are no general existence results like theorem \ref{Horm}.
For non linear second order equations a good reference is
\cite{Gilbarg} and for non linear systems \cite{Giaquinta93} and
\cite{giaquinta83}. A related issue that was not discussed here is
regularity. We have assumed that the functions and the boundary are
smooth.  Regularity properties are crucial for non linear systems, see
\cite{Gilbarg}, \cite{Giaquinta93} and \cite{giaquinta83} and
references there.

In section \ref{sec:so} we have followed \cite{Folland}. Good
references for this section are also \cite{Evans98}, \cite{Gilbarg}
and \cite{McOwen96}. For section \ref{sec:es}, an introductory book is
\cite{Renardy04}; more advanced material can be found in
\cite{Taylor96} and \cite{Agranovich97}; for a complete discussion see \cite{Hoermander85},
\cite{Hoermander63} and \cite{Wloka95}.

\section*{Acknowledgments}
 Part of the material of these notes was presented in two lectures at
 the school on ``Structure and dynamics of compact objects'' 
 which took place in the Albert Einstein Institut on 20-25/09,
 2004; as a part of the SFB/TR7 project.  
 I would like to thank the organizers S. Boutloukos, J. Frauendiener
 and S.  Husa for the invitation. I would like also to thank F. Beyer  for his
careful reading of the manuscript.

 This work has been supported by the Sonderforschungsbereich SFB/TR 7
 of the Deutsche Forschungsgemeinschaft.


\begin{thebibliography}{10}

\bibitem{Agmon59}
S.~Agmon, A.~Douglis, and L.~Nirenberg.
\newblock Estimates near the boundary for solutions of elliptic partial
  differential equations satisfying general boundary conditions {I}.
\newblock {\em Comm. Pure App. Math.}, 12:623--727, 1959.

\bibitem{Agmon64}
S.~Agmon, A.~Douglis, and L.~Nirenberg.
\newblock Estimates near the boundary for solutions of elliptic partial
  differential equations satisfying general boundary conditions. {II}.
\newblock {\em Comm. Pure App. Math.}, 17:35--92, 1964.

\bibitem{Agranovich97}
M.~S. Agranovich.
\newblock Elliptic boundary problems.
\newblock In {\em Partial differential equations, IX}, volume~79 of {\em
  Encyclopaedia Math. Sci.}, pages 1--144. Springer, Berlin, 1997.
\newblock Translated from the Russian by the author.

\bibitem{Bartnik:2002cw}
R.~Bartnik and J.~Isenberg.
\newblock The constraint equations.
\newblock In P.~T. Chru\'sciel and H.~Friedrich, editors, {\em The {E}instein
  equations and large scale behavior of gravitational fields}, pages 1--38.
  Birhäuser Verlag, Basel Boston Berlin, 2004, gr-qc/0405092.

\bibitem{Browder59}
F.~E. Browder.
\newblock Estimates and existence theorems for elliptic boundary value
  problems.
\newblock {\em Proc. Nat. Acad. Sci. U.S.A.}, 45:365--372, 1959.

\bibitem{Chrusciel:2003sr}
P.~T. Chru{\'s}ciel and E.~Delay.
\newblock On mapping properties of the general relativistic constraints
  operator in weighted function spaces, with applications.
\newblock 2003, gr-qc/0301073.

\bibitem{Corvino:2003sp}
J.~Corvino and R.~M. Schoen.
\newblock On the asymptotics for the vacuum {E}instein constraint equations.
\newblock 2003, gr-qc/0301071.

\bibitem{Dain03}
S.~Dain.
\newblock Trapped surfaces as boundaries for the constraint equations.
\newblock {\em Class. Quantum. Grav.}, 21(2):555--573, 2004, gr-qc/0308009.

\bibitem{Douglis55}
A.~Douglis and L.~Nirenberg.
\newblock Interior estimates for elliptic systems of partial differential
  equations.
\newblock {\em Comm. Pure App. Math.}, 8:503--538, 1955.

\bibitem{Evans98}
L.~C. Evans.
\newblock {\em Partial differential equations}, volume~19 of {\em Graduate
  Studies in Mathematics}.
\newblock American Mathematical Society, Providence, RI, 1998.

\bibitem{Folland}
G.~B. Folland.
\newblock {\em Introduction to Partial Differential Equation}.
\newblock Princeton University Press, Princeton, {NY}, 1995.

\bibitem{giaquinta83}
M.~Giaquinta.
\newblock {\em Multiple Integrals in the Calculus of Variations and Nonlinear
  Elliptic Systems}, volume 105 of {\em Annals of Mathematics Studies}.
\newblock Princeton University Press, Princeton, New Jersey, 1983.

\bibitem{Giaquinta93}
M.~Giaquinta.
\newblock {\em Introduction to Regularity Theory for Nonlinear Elliptic
  Systems}.
\newblock Lectures in Mathematics (ETH Zürich). Birkh{\"a}user Verlag, Berlin,
  1993.

\bibitem{Gilbarg}
D.~Gilbarg and N.~S. Trudinger.
\newblock {\em Elliptic Partial Differential Equations of Second Order}.
\newblock Springer-Verlag, Berlin, 1983.

\bibitem{Hoermander63}
L.~H{\"o}rmander.
\newblock {\em Linear partial differential operators}, volume 116 of {\em Die
  Grundlehren der mathematischen Wissenschaften}.
\newblock Academic Press Inc., Publishers, New York, 1963.

\bibitem{Hoermander85}
L.~H{\"o}rmander.
\newblock {\em The analysis of linear partial differential operators. {III}},
  volume 274 of {\em Grundlehren der Mathematischen Wissenschaften}.
\newblock Springer-Verlag, Berlin, 1985.
\newblock Pseudodifferential operators.

\bibitem{Hoermander85b}
L.~H{\"o}rmander.
\newblock {\em The analysis of linear partial differential operators. {IV}},
  volume 275 of {\em Grundlehren der Mathematischen Wissenschaften}.
\newblock Springer-Verlag, Berlin, 1985.
\newblock Fourier integral operators.

\bibitem{Kato80}
T.~Kato.
\newblock {\em Perturbation Theory for Linear Operators}, volume 132 of {\em
  Grundlehren der mathematischen Wissenschaften}.
\newblock Springer-Verlag, 1980.

\bibitem{Lions72}
J.~L. Lions and E.~Magenes.
\newblock {\em Non-homogeneous boundary value problems and applications.},
  volume 181 of {\em Grundlehren der mathematischen Wissenschaften}.
\newblock Springer-Verlag, New York, 1972.
\newblock Translated from the French by P. Kenneth.

\bibitem{Marsden83}
J.~E. Marsden and T.~J. Hughes.
\newblock {\em Mathematical Foundations of Elasticity}.
\newblock Prentice-Hall civil engineering and engineering mechanics series.
  Prentice Hall, Inc., Englewood Cliffs, New Jersey, 1983.

\bibitem{Maxwell03}
D.~Maxwell.
\newblock Solutions of the {E}instein constraint equations with apparent
  horizon boundary.
\newblock 2003, gr-qc/0307117.

\bibitem{McOwen96}
R.~C. McOwen.
\newblock {\em Partial Differential Equation}.
\newblock Prentice Hall, New Jersey, 1996.

\bibitem{Miao03}
P.~Miao.
\newblock On existence of static metric extensions in general relativity.
\newblock {\em Commun. Math. Phys.}, 241(1):27--46, 2003.

\bibitem{Nirenberg55}
L.~Nirenberg.
\newblock Remarks on strongly elliptic partial differential equations.
\newblock {\em Comm. Pure App. Math.}, 8:649--675, 1955.

\bibitem{Nirenberg56}
L.~Nirenberg.
\newblock Estimates and existence of solutions of elliptic equations.
\newblock {\em Comm. Pure App. Math.}, 9:509--529, 1956.

\bibitem{Parker82}
T.~Parker and C.~H. Taubes.
\newblock On {W}itten's proof of the positive energy theorem.
\newblock {\em Comm. Math. Phys.}, 84(2):223--238, 1982.

\bibitem{Popivanov97}
P.~R. Popivanov and D.~K. Palagachev.
\newblock {\em The degenerate oblique derivative problem for elliptic and
  parabolic equations}, volume~93 of {\em Mathematical Research}.
\newblock Akademie Verlag, Berlin, 1997.

\bibitem{Renardy04}
M.~Renardy and R.~C. Rogers.
\newblock {\em An introduction to partial differential equations}, volume~13 of
  {\em Texts in Applied Mathematics}.
\newblock Springer-Verlag, New York, second edition, 2004.

\bibitem{MR83e:83036}
O.~Reula.
\newblock Existence theorem for solutions of {W}itten's equation and
  nonnegativity of total mass.
\newblock {\em J. Math. Phys.}, 23(5):810--814, 1982.

\bibitem{Reula89}
O.~Reula.
\newblock On existence and behaviour of asymptotically flat solutions to the
  stationary {E}instein equations.
\newblock {\em Commun. Math. Phys.}, 122:615--624, 1989.

\bibitem{reula:84}
O.~Reula and K.~P. Tod.
\newblock Positivity of the {B}ondi energy.
\newblock {\em J. Math. Phys.}, 25(4):1004--1008, 1984.

\bibitem{Schechter59}
M.~Schechter.
\newblock General boundary value problems for elliptic partial differential
  equations.
\newblock {\em Comm. Pure Appl. Math.}, 12:457--486, 1959.

\bibitem{Smarr78a}
L.~Smarr and J.~York.
\newblock Radiation gauge in general relativity.
\newblock {\em Phys. Rev. D}, 17:1945, 1978.

\bibitem{Sohr01}
H.~Sohr.
\newblock {\em The {N}avier-{S}tokes equations}.
\newblock Birkh\"auser Advanced Texts. Birkh\"auser Verlag, Basel, 2001.
\newblock An elementary functional analytic approach.

\bibitem{Taylor96}
M.~E. Taylor.
\newblock {\em Partial differential equations. {I}}, volume 115 of {\em Applied
  Mathematical Sciences}.
\newblock Springer-Verlag, New York, 1996.

\bibitem{Thompson69}
J.~L. Thompson.
\newblock Some existence theorems for the traction boundary value problem of
  linearized elastostatics.
\newblock {\em Arch. Rational Mech. Anal.}, 32:369--399, 1969.

\bibitem{Witten81}
E.~Witten.
\newblock A new proof of the positive energy theorem.
\newblock {\em Comm. Math. Phys.}, 80(3):381--402, 1981.

\bibitem{Wloka95}
J.~T. Wloka, B.~Rowley, and B.~Lawruk.
\newblock {\em Boundary value problems for elliptic systems}.
\newblock Cambridge University Press, Cambridge, 1995.

\bibitem{York73}
J.~W. York, Jr.
\newblock Conformally invariant orthogonal decomposition of symmetric tensor on
  {R}iemannian manifolds and the initial-value problem of general relativity.
\newblock {\em J. Math. Phys.}, 14(4):456--464, 1973.

\end{thebibliography}

\end{document}